\documentclass[10pt,reprint]{revtex4-1}

\usepackage{iopams} 
\usepackage{amsmath}
\usepackage{graphicx}
\usepackage[utf8]{inputenc}
\usepackage[T1]{fontenc}
\usepackage{mathptmx}
\usepackage{caption}
\usepackage{subcaption}
\usepackage{bm}

\begin{document}

\title{A Path Integral approach to Quantum Fluid Dynamics}

\author{Sagnik Ghosh}
\address{ 
Indian Institute of Science Education and Research, Pune-411008, India}
%\ead{submissions@iop.org}

\author{Swapan K Ghosh}
% \homepage{http://www.Second.institution.edu/~Charlie.Author.}
\address{%
UM-DAE Centre for Excellence in Basic Sciences, University of Mumbai, Kalina, Santacruz, Mumbai-400098, India 
}%

\email{swapan.ghosh@cbs.ac.in}

%\vspace{10pt}

\date{\today}

\begin{abstract}
In this work we develop an alternative approach for solution of Quantum Trajectories using the Path Integral method. The state-of-the-art technique in the field is to solve a set of non-linear, coupled partial differential equations (PDEs) simultaneously. We opt for a fundamentally different route. We first derive a general closed form expression for the Path Integral propagator valid for any general potential as a functional of the corresponding classical path. The method is exact and is applicable in many dimensions as well as multi-particle cases. This, then, is used to compute the Quantum Potential (QP), which, in turn, can generate the Quantum Trajectories. For cases, where closed form solution is not possible, the problem is formally boiled down to solving the classical path as a boundary value problem. The work formally bridges the Path Integral approach with Quantum Fluid Dynamics. As a model application to illustrate the method, we work out a toy model viz. the double-well potential, where the boundary value problem for the classical path has been computed perturbatively, but the Quantum part is left exact. Using this we delve into seeking insight in one of the long standing debates with regard to Quantum Tunneling.
\\
\textbf{Submitted to:} \textit{J. Phys. A: Math. Theor.}
\end{abstract}

%\noindent{Keywords}: Path Integrals, Quantum Fluid Dynamics.
%\maketitle
\maketitle

\section{Introduction}

    Quantum Fluid Dynamics (QFD) \cite{De_Broglie,Madelung,Bohm1,Bohm2,Takabayasi} is a tool for the study of a quantum system as motion of the probability fluid. It is widely employed in theoretical chemistry, condensed matter physics and material sciences. The main advantage of this approach starts with recasting of the Schrodinger's equation into two coupled differential equations involving real quantities, formally known as the Madelung transformation\cite{Madelung}. This in particular often make it computationally more useful over the usual formulation in the study of dynamics of various molecular or quantum optics systems.

    \par The usual route for computation in QFD is in some or the other way related to solving a set of coupled differential equations simultaneously\cite{Wyattbook}. In this study, we seek to develop a fundamentally different route to achieve the quantum trajectories by invoking the Path Integral method, which reduces the problem of finding the quantum trajectories to solving the corresponding classical trajectory as a boundary value problem (BVP). We show that, the QP can be expressed in terms of a series, each term of which can be computed in a closed form as a functional of the classical path, $x^{cl}(t)$.  These analytical functional forms can be provided directly as an input to the machine simplifying the algorithm for computation of Quantum Trajectories to computation of a BVP. This method, hence, is expected to be computationally superior to the Quantum Trajectory Method (QTM)\cite{Wyatt1} due to Wyatt et. al, which sets the current standard for time complexity.
    
    \par Our derivation sets off from the Feynman theory of Path Integrals.\cite{Dirac,Feynman}.\footnote{\cite{LNM523} covers the detail nitty-gritties of the mathematical foundations of Feynman Path Integrals. For the physical theory \cite{PIQM} is the standard introduction.} Firstly, we develop a general solution for the propagator. We derive each term in the series as a closed form functional of the classical path, which then can be used to obtain the Quantum Trajectories.  The method is similar to Ursell decomposition of statistical mechanical partition functions.

    \par The paper is structured as follows. We first discuss in short, the results of Path Integrals and QFD, relevant to our work. Then we present our derivation, firstly for 1-D single particle case, and then generalize it to higher dimensions and many-body systems. Then we present a toy example of an 1-D double well potential, solved analytically using the ideas we develop for the purpose of illustration of the method. The classical boundary value problem, in this case, has been obtained perturbatively using Linsted-Poincare method. We show that the solution obtained by our method exhibits tunneling even in the very first order of perturbation of the classical path, when the quantum part of the calculation is left exact. We conclude indicating some applications and further directions of study.
    
\section{Formultion: A General Solution for Path Integral Propagator}

    In this section we present an exact closed form series solution for the Kernel of the Path Integral (the two-point correlator) for a quantum particle for any general applied potential $V(x)$. In Path Integral formalism, the Kernel can be solved exactly for some simple systems viz the free particle, Simple Harmonic Oscillator and approximately for a few more complicated systems. To the best of our knowledge, the solution presented here is the first exact solution without assuming any special property of the applied potential other than continuity and differentiabilty. We express the Kernel as a functional of the classical path.
    
    The correlator gives the amplitude at a point $x_t$, after a time of flight $t$, given that it has started from an initial point $x_0$. It is also the Green's function to the Schrodinger equation \cite{Feynman}, rendering it the formal equivalence with usual formulations. Equipped with this, we pose the solution to quantum trajectories, in an integral equation framework.  
    
    \par According to Path Integral approach, the dynamics of a quantum particle from one point to another, is contributed by all the possible paths that connects them. The contribution from each path is weighed by $\exp{(iS[x(t)]/\hbar)}$, where $S[x(t)]$ is the corresponding Action or Hamilton's principal function. The net propagator $K(x_t,t;x_0,0)$ is obtained by adding the contributions from all paths. The integral
    
     \begin{align}\label{dynamics}
         \psi(x_t,t)=\int^\infty_{-\infty}K(x_t,t;x_0,0)\psi_0(x_0)dx_0
     \end{align}
    
    gives the time evolution of the wave-function.

    \par Next, we exploit the Madelung transformation\cite{Madelung}, to express $\psi(t)$ in polar form. \footnote{In the language of QFD, $S_M$ is also known as action. We denote this with subscript $M$ (after Madelung) to differentiate it form the action in Path Integral, which is used with no subscript. These are different quantities and should not be confused with each other. For e.g $S_M$ has contribution from the initial distribution, whereas $S$ in Path Integral is generic}. The Madelung Transformation provides a way of looking at the quantum systems as probability fluids, by decoupling the Schrodinger equation into the following coupled equations. \footnote{Later more general derivation has been carried out by Takabyashi\cite{Takabayasi}} 
    
    For a particle of mass M in a potential $V(x)$ we have,
    
    \begin{align}\label{cont}
        \nabla(\rho \mathbf{v})+ \frac{\partial\rho}{\partial t}=0
    \end{align}
    
    \begin{align}\label{jacobian}
     -\frac{\partial S}{\partial t}= \frac{1}{2}M\mathbf{v}^2 + V(x)+ Q(x,t) 
    \end{align}
    
    Here $\rho(x,t)=R^2(x,t)$, $\mathbf{v}(x,t)=\nabla\ S(x,t)/M$ and the quantum potential, Q(x,t) is given by,
    
    \begin{align}\label{Q}
        Q(x,t)=-\frac{\hbar ^2}{2M}\frac{\nabla ^2 R(x,t)}{R(x,t)}
    \end{align}

    In this work we look for footing of the QP in the Feynman Path Integrals, with the aim of developing formal functional expression of the former in terms of the classical path and the Cauchy data. We first treat the one dimensional system and then extend it to higher dimensions.

    \par Let's denote the classical path of the system, from a initial position $x_0$, to a final point $x_t$ after a time of flight t, by $x^{cl}(t)$. Being solution to a second order Boundary Value Problem, it parametrically depends on, $x_0$ and $x_t$.

    A general path of the system, denoted by $x(t)$, is any continuous function of $t$ that is constrained by, $x(0)=x_0$ and $ x(t)=x_t$. Any such general path can be broken down as, 
    
     \begin{align}\label{disc}
         x(t)=x^{cl}(t)+y(t).
     \end{align}
    
     varying $y(t)$ to span the space of continuous functions on $[t_0,t]$ subject to the boundary conditions, $y(t_0)=y(t)=0$ tantamounts to span the whole space of allowed paths of Path Integral formulation, over which we need to sum over to obtain the propagator.\cite{PIQM}

     \par One way to compute the Path Integral is to first discretise the time axis into n moments, separated by $\epsilon$ units. 
     
        \begin{align*}
             &t_{j}=j\epsilon, \; \\&  t_{0}=0 \;, \;\;t_{n}=t
        \end{align*}
     
     The Path Integral can then be recovered in the limit $n\rightarrow 0, \epsilon\rightarrow \infty$, such that the product remains constant, $n\epsilon=t$, after the n-dimensional integral is carried out. Equation (\ref{disc}) under the discretisation becomes, 
     
        \begin{align}\label{Split}
             x(t_j)=x_j=x^{cl}_j+y_j.
        \end{align}
     
     at each of these moments. By definition $y_0=y_n=0$. 
     \par The propagator is defined as,

         \begin{align}\label{Propagator_defn}
            K(x_t,t;x_0,0)= \int \exp\big[\;\frac{i}{\hbar}\int^t_0 \mathcal{L}(x(t'),\dot{x}(t'),t')\;dt'\big]\;\; \mathcal{D}x
        \end{align}

     where $\mathcal{L}(x,\dot{x},t')$ is the classical Lagrangian of the system. In the discretised version it is rephrased  as,

        \begin{widetext}
            \begin{align}\label{start}
            \lim_{n \rightarrow\infty} \; \frac{1}{A} \int^{\infty}_{-\infty}\cdots\int^{\infty}_{-\infty}\exp\Big( \frac{i\epsilon}{\hbar} \sum_{j=1}^n \Big[\frac{M}{2\epsilon^2}(x_j-x_{j-1})^2-V(x_j)\Big]\Big) dx_1\cdots dx_n 
            \end{align}
        \end{widetext}
    
    where $A$ is the normalization constant which, in general, depends on $n$. Throughout the paper this limit is understood to be taken as  $n \rightarrow \infty$ and $\epsilon \rightarrow 0$ simultaneously, with $n\epsilon=t$ remaining constant. 
    
    \par The potential over the general path can be Taylor expanded the around the classical position at each moment, as \footnote{The radius of convergence depends on the V(x), its higher derivatives and the convergence is not guaranteed in general.}
    
    \begin{align}\label{TaylorExpansion}
        V(x_j)=V(x^{cl}_j+y_j)=V(x^{cl}_j)+\sum_{m=1}^{\infty}\frac{y^m_j}{m!}\frac{\partial^m}{\partial x^m}V(x)|_{x=x^{cl}_j}
    \end{align}

    Substituting (\ref{Split}),(\ref{TaylorExpansion}) into (\ref{start}) we obtain the propagator as,
    
        \begin{widetext}
            \begin{multline}\label{GeneralIntegral}
            \lim_{n \rightarrow\infty} \; \frac{1}{A} \int^{\infty}_{-\infty}\cdots\int^{\infty}_{-\infty} \exp\Big( \frac{i\epsilon}{\hbar}\sum_{j=1}^n \Big[\frac{M}{2\epsilon^2}(x^{cl}_j-x^{cl}_{j-1})^2 + 
            \frac{M}{2\epsilon^2}(x^{cl}_j-x^{cl}_{j-1})(y_j-y_{j-1})+\frac{M}{2\epsilon^2}  (y_j-y_{j-1})^2 \\
                -V(x^{cl}_j)-\sum_{m=1}^{\infty} \frac{y_j^m}{m!}  \frac{\partial^m}{\partial x^m}V(x)|_{x=x^{cl}_j}\Big]\Big) dy_1\cdots dy_n
            \end{multline}
        \end{widetext}

    The terms in (\ref{GeneralIntegral}), that are independent of $y_j$, can be taken out of the integrals. They add upto $\exp{\big[i S^{cl}(x_t,t;x_0,0)/\hbar\big]}$, where $S^{cl}(x_t,t;x_0,0)$ is the Classical Action. The linear terms in  $y_j$ do not contribute \cite{PIQM}.

    \par The primary complication in the generalization from the corresponding case of the Harmonic Oscillator, that forms the thesis of this work, stems form the non-vanishing higher derivatives of the potential ($m=3$ and higher), which in general depends on the positional variables $x^{cl}_j$. These show up in the pre-factor, and makes it position dependent. 
    
    \par The rest of the general integral, that denotes the pure quantum part, is

        \begin{widetext}
            \begin{align}\label{GeneralIntegral2}    
            \lim_{n \rightarrow\infty}\frac{1}{A} \int^{\infty}_{-\infty}\cdots\int^{\infty}_{-\infty} \exp\Big( \frac{i\epsilon}{\hbar}\sum_{j=1}^n \Big[\frac{M}{2\epsilon^2} (y_j-y_{j-1})^2 - \frac{y_j^2}{2!}\frac{\partial^2}{\partial x^2}V(x)|_{x=x^{cl}_j} -\sum_{m=3}^{\infty} \frac{y_j^m}{m!} \frac{\partial^m}{\partial x^m}V(x)|_{x=x^{cl}_j}\Big]\Big) dy_1\cdots dy_n
            \end{align}
        \end{widetext}

    \par The main sketch of our construction follows from the fact that the $n$ dimensional integral can be reduced to combination of simpler integrals which can be computed in closed form. First we systematically expand the integral, substitute for the simpler integrals, and then wind it up back. Convergence conditions are not discussed. 
    
    \par We do a series of rearrangements to (\ref{GeneralIntegral2}) and obtain,

        \begin{widetext}
    
            \begin{multline}
            \lim_{n \rightarrow\infty}\frac{1}{A} \int^{\infty}_{-\infty}\cdots\int^{\infty}_{-\infty} \exp \big(\frac{i\epsilon}{\hbar}\sum_{j'=1}^n \Big[\frac{M}{2\epsilon^2} (y_{j'}-y_{j'-1})^2  -\frac{y_{j'}^2}{2!} \frac{\partial^2}{\partial x^2}V(x)|_{x=x^{cl}_{j'}}\Big]\big)
            \\ \times  \sum_{k=0}^{\infty}\frac{1}{k!}\Bigg(\frac{-i\epsilon}{\hbar}\sum_{j=1}^n\sum_{m=3}^{\infty} \frac{y_j^m}{m!} \frac{\partial^m}{\partial x^m}V(x)|_{x=x^{cl}_j}\Bigg)^k \;dy_1\cdots dy_n  
            \end{multline}
            
        Since the limit is independent of $k$, we are free to take the $k$ sum out. 
    
            \begin{multline}
            \sum_{k=0}^{\infty}\lim_{n \rightarrow\infty}\frac{1}{A} \int^{\infty}_{-\infty}\cdots\int^{\infty}_{-\infty} \exp \big(\frac{i\epsilon}{\hbar}\sum_{j'=1}^n \Big[\frac{M}{2\epsilon^2} (y_{j'}-y_{j'-1})^2  -\frac{y_{j'}^2}{2!} \frac{\partial^2}{\partial x^2}V(x)|_{x=x^{cl}_{j'}}\Big]\big)
            \\ \times \Big[ \frac{1}{k!}\Bigg(\frac{-i\epsilon}{\hbar}\sum_{j=1}^n\sum_{m=3}^{\infty} \frac{y_j^m}{m!} \frac{\partial^m}{\partial x^m}V(x)|_{x=x^{cl}_j}\Bigg)^k\Big] dy_1\cdots dy_n
            \end{multline}

        Which can be expressed as,
  
            \begin{multline}\label{GeneralIntegral3}
            \sum_{k=1}^{\infty}\frac{1}{k!}
            \; \lim_{n \rightarrow\infty} \; \frac{1}{A} \Big(\frac{-i\epsilon}{\hbar}\Big)^k \int^{\infty}_{-\infty}\cdots\int^{\infty}_{-\infty} \exp\big(\frac{i\epsilon}{\hbar}\sum_{j'=1}^n \Big[\frac{M}{2\epsilon^2} (y_{j'}-y_{j'-1})^2  -\frac{y_{j'}^2}{2!} \frac{\partial^2}{\partial x^2}V(x)|_{x=x^{cl}_{j'}}\Big]\big)
            \\ \times \sum_{m_1=3}^{\infty}\sum_{m_2=3}^{\infty}\cdots\sum_{m_k=3}^{\infty} \sum_{j_1=1}^n\sum_{j_2=1}^n \cdots \sum_{j_k=1}^n
            \Big[\prod_{\alpha=1}^{k} \frac{1}{m_{\alpha}!} \frac{\partial^{m_{\alpha}}}{\partial x^{m_{\alpha}}}V(x)|_{x=x^{cl}_{j_{\alpha}}} \;\;y_{j_{\alpha}}^{m_{\alpha}} \Big] dy_1\cdots dy_n
            \end{multline}
            
    \pagebreak
        \end{widetext}

        Expression (\ref{GeneralIntegral3}) suggests that the individual n-dimensional integrals that we are dealing with are basic Gaussian integrals that can be computed in closed form. From the symmetry of the integral, it survives if and only if all $m_{\alpha}$ are even.

         Since the limit does not depend on $m$, the multi-dimensional sum over $m$ can be taken out. 
  \begin{widetext}
            \begin{multline}\label{Expression}
            \sum_{k=0}^{\infty}\frac{1}{k!} \Big(\frac{-i}{\hbar}\Big)^k\sum_{m_1=3}^{\infty}\cdots\sum_{m_k=3}^{\infty} 
            \lim_{n \rightarrow\infty} \frac{1}{A} \epsilon^k \sum_{j_1=1}^n\cdots \sum_{j_k=1}^n \int^{\infty}_{-\infty}\cdots\int^{\infty}_{-\infty}
            \Big[\prod_{\alpha=1}^{k} \frac{1}{m_{\alpha}!} \frac{\partial^{m_{\alpha}}}{\partial x^{m_{\alpha}}}V(x)|_{x=x^{cl}_{j_{\alpha}}} y_{j_{\alpha}}^{m_{\alpha}} \Big] 
            \\ \times \exp\big(\frac{i\epsilon}{\hbar}\sum_{j'=1}^n \Big[\frac{M}{2\epsilon^2} (y_{j'}-y_{j'-1})^2  -\frac{y_{j'}^2}{2!} \frac{\partial^2}{\partial x^2}V(x)|_{x=x^{cl}_{j'}}\Big]\big)
            dy_1\cdots dy_n
            \end{multline}
  \end{widetext}
        \par We define the instantaneous angular frequency $\omega_j$ as,
         
         \begin{align}
            \omega_j^2= \frac{1}{M}\frac{\partial^2}{\partial x^2}V(x)|_{x=x^{cl}_j}
         \end{align}
        
        In terms of this Eq. (\ref{Expression}) becomes,
        
        \begin{widetext}
            \begin{multline}\label{GeneralIntegral4}
            \sum_{k=0}^{\infty}\frac{1}{k!} \Big(\frac{-i}{\hbar}\Big)^k\sum_{m_1=3}^{\infty}\cdots\sum_{m_k=3}^{\infty} 
            \lim_{n \rightarrow\infty} \frac{1}{A} \epsilon^k \sum_{j_1=1}^n\cdots \sum_{j_k=1}^n \int^{\infty}_{-\infty}\cdots\int^{\infty}_{-\infty}
            \Big[\prod_{\alpha=1}^{k} \frac{1}{m_{\alpha}!} \frac{\partial^{m_{\alpha}}}{\partial x^{m_{\alpha}}}V(x)|_{x=x^{cl}_{j_{\alpha}}} y_{j_{\alpha}}^{m_{\alpha}} \Big]\\
            \exp\big(\frac{i M}{2\epsilon\hbar}\sum_{j'=1}^n \Big[ (y_{j'}-y_{j'-1})^2  -\epsilon^2\omega_{j'}^2y_{j'}^2 \Big]\big)dy_1\cdots dy_n
            \end{multline}
        
        \end{widetext}
        
        \par The integrals can be computed as appropriate partial derivatives w.r.t various coefficients of $y_j^2$ of the following integral,

        \begin{align}
            \frac{1}{A}\int^{\infty}_{-\infty}\int^{\infty}_{-\infty} \exp \big(\frac{i\epsilon}{\hbar}\sum_{j=1}^n\Big[\frac{M}{2\epsilon^2} (y_j-y_{j-1})^2 -\frac{M}{2}\omega_j^2 y_j^2 \Big]\big)dy_1\cdots dy_n
        \end{align}
        
        which itself evaluates to,
        
        \begin{align}\label{determinant}
            \Big(\frac{M}{2\pi i \hbar \epsilon}\Big)^{\frac{n+1}{2}}\Big(\frac{2\pi i \hbar}{M \epsilon}\Big)^{\frac{n}{2}}\Big(\frac{\sin[\epsilon\sum_{j=1}^{n-1}\omega_j]}{\epsilon\sum_{j=1}^{n-1}\omega_j}\Big)^{-\frac{1}{2}}
        \end{align}
        
        This result follows from spectral theory of tri-digonal matrices \cite{Tridiagonal1}, and is exact in the $n \rightarrow\infty$ limit. The individual integrals are,
    
    \begin{widetext}
    \begin{multline}
            \int^{\infty}_{-\infty}\cdots\int^{\infty}_{-\infty}\exp\big(\frac{i\epsilon}{\hbar}\sum_{j'=1}^n \Big[\frac{M}{2\epsilon^2} (y_{j'}-y_{j'-1})^2  -\frac{M \omega_{j'}^2 y_{j'}^2}{2} \Big]\big)
            \Big[\prod_{\alpha=1}^k y_{j_{\alpha}}^{m'_{\alpha}} \Big]
            dy_1\cdots dy_n\\
            =\sqrt{\frac{M }{2\pi i \hbar t }}\prod_{\alpha=1}^k\Big[\frac{\hbar}{-i M  \omega_j}\frac{\partial}{\partial \phi}\Big]^{m'_{\alpha}}\sqrt{\frac{\phi}{\sin{(\phi)}}}
    \end{multline}
    
    where $m'=m/2$ for all $\alpha$ and we define $\phi$ as,
        
        \begin{align}
            \phi=\epsilon\sum_{j=1}^{n-1}\omega_j=\int_0^t \omega(t') dt'
        \end{align}

    Substituting this back to (\ref{Expression}) we obtain the quantum correction as,
 
    \begin{multline}
        \sqrt{\frac{M }{2\pi i \hbar t }}\sum_{k=0}^{\infty}\frac{1}{k!} \Big(\frac{-i}{\hbar}\Big)^k\sum_{m_1=3}^{\infty}\cdots\sum_{m_k=3}^{\infty} \frac{\partial}{\partial \phi}^{\sum_{\alpha} m'_{\alpha}}\sqrt{\frac{\phi}{\sin{(\phi)}}} 
            \lim_{n \rightarrow\infty}  \epsilon^k \times 
            \sum_{j_1=1}^n\cdots \sum_{j_k=1}^n \Big[\prod_{\alpha=1}^{k} \frac{1}{m_{\alpha}!} \Big(\frac{\hbar}{-i M  \omega_{j_{\alpha}}}\Big)^{m'_{\alpha}}\frac{\partial^{m_{\alpha}}}{\partial x^{m_{\alpha}}}V(x)|_{x=x^{cl}_{j_{\alpha}}} \Big] 
    \end{multline}
    \end{widetext}

    At the limit the $j$ sums becomes integrals,
    
    \begin{widetext}
    \begin{multline}
        \sqrt{\frac{M }{2\pi i \hbar t }}\sum_{k=0}^{\infty}\frac{1}{k!} \Big(\frac{-i}{\hbar}\Big)^k\sum_{m_1=3}^{\infty}\cdots\sum_{m_k=3}^{\infty} \prod_{\alpha=1}^{k} \frac{1}{m_{\alpha}!} \Big(\frac{\hbar}{-i M }\Big)^N \frac{\partial}{\partial \phi}^N \sqrt{\frac{\phi}{\sin{(\phi)}}} 
             \Big[  \prod_{\alpha=1}^{k} \int_0^t \frac{1}{\omega(t_{\alpha})^{m'_{\alpha}}}\frac{\partial^{m_{\alpha}}}{\partial x^{m_{\alpha}}}V(x)|_{x=x^{cl}({t_{\alpha}})}dt_{\alpha} \Big] 
    \end{multline}
 
    \end{widetext}

    where, $N=\sum_{\alpha} m'_{\alpha}$.   
    
    This series expresses the complete Path Integral as a functional of the classical path. The complete expression of the Kernel is,
    
    \begin{widetext}
    \begin{multline}\label{FinalExpression}
        K(x_b,x_a; t, 0 )=\\
        \sqrt{\frac{M }{2\pi i \hbar t }} \exp{\frac{i S^{cl}(x_b,x_a; t)}{\hbar}}\sum_{k=0}^{\infty}\frac{1}{k!} \Big(\frac{-i}{\hbar}\Big)^k\sum_{m_1=3}^{\infty}\cdots\sum_{m_k=3}^{\infty} \prod_{\alpha=1}^{k} \frac{1}{m_{\alpha}!} \Big(\frac{\hbar}{-i M }\Big)^N \frac{\partial}{\partial \phi}^N \sqrt{\frac{\phi}{\sin{(\phi)}}} 
             \Big[  \prod_{\alpha=1}^{k} \int_0^t \frac{1}{\omega(t_{\alpha})^{m'_{\alpha}}}\frac{\partial^{m_{\alpha}}}{\partial x^{m_{\alpha}}}V(x)|_{x=x^{cl}({t_{\alpha}})}dt_{\alpha} \Big]
    \end{multline}
    \end{widetext}
    
    Equipped with (\ref{FinalExpression}) the dynamics from an initial wave-function can be obtained as,
    
     \begin{align}
         \psi(x_t,t)=&\int^\infty_{-\infty}K(x_t,t;x_0,0)\psi_0(x_0)dx_a
     \end{align}
    
    which then, after a Madelung transformation, can be used to obtain the Quantum Potential and solve for the quantum Trajectories. Expression (\ref{FinalExpression}) extends the domain of analytically solvable Path Integrals to any system whose classical action (and the path) can be computed in a closed form. In cases where the trajectories cannot be obtained in closed form, we have argued that the problem can be reduced to numerically computing the classical path as a BVP. 
    
    The following section discusses the generalisation to multidimensional and multi particle cases.

\section{Multidimensional Generalisation}

     Most of the derivation goes through while generalising it to higher dimensions. The only change is brought into via the Taylor Expansion, which in this case becomes multidimensional. The generalisation to multidimensional (and multi-particle) cases are done by perceiving it in the generalised coordinates.
     
     Consider a three dimensional space with, $\mathbf{x}=(x_1,x_2,x_3)$. The potential remains a scalar function of path variables, $V(\mathbf{x})$. $\mathbf{x}(t)$ defines a path in this three dimensional space. Similar to (\ref{disc}), we can split such a general path as,
    \begin{align}
         \mathbf{x}(t)=\mathbf{x}^{cl}(t)+\mathbf{y}(t).
     \end{align}
     This brings in changes in the Taylor expansion (\ref{TaylorExpansion})
     
     \begin{align}
        V(\mathbf{x}_j)=V(\mathbf{x}^{cl}_j+\mathbf{y}_j)=V(\mathbf{x}^{cl}_j)+\sum_{m=1}^{\infty}\frac{\mathbf{y}^m_j}{m!}\mathbf{D}^m V(\mathbf{x})|_{\mathbf{x}=\mathbf{x}^{cl}_j}
    \end{align}
    
    where,
    
    \begin{align}
     \mathbf{D}^m V(\mathbf{x})= \frac{\partial^m V(\mathbf{x})}{\partial x_1^{\alpha_1}\partial x_2^{\alpha_2} \partial x_3^{\alpha_3}}
    \\\alpha_1+\alpha_2+\alpha_3=m
    \end{align}
    
    Accordingly the integrals over each $dy_j$ in (\ref{GeneralIntegral}) is replaced by the corresponding volume integral $\mathbf{dy}_j$. In essence the higher dimensional case is like a foliation of space of the one dimensional case, where each one dimensional line integral foliates into a higher dimensional volume. For the multi-particle case the one-dimensional case foliates into a $3n$ dimensional space. The rest of the derivation goes through exactly. The $\mathbf{y}$ independent parts generate the exponentiation of classical action, and the prefactor is written as a series, of various moments of $k$, only in this case the partial derivatives are more diverse.

     The formula (\ref{FinalExpression}) provides the most general formal expression for the Path Integral propagator in closed form, subject to the availability of the classical path in a closed form expression. In the following section we illustrate this method by solving for a toy example and delve into Quantum Tunneling.

\section{An Example: Double Well Potential}

     \par As a model application, in this section we calculate the quantities for a double well potential. The standard treatment of this system can be found in many textbooks.\cite{LandauQuantum,Schulman}. Despite its simplicity, it has wide applcation in condensed matter theory\cite{PhysRevA.Anharmonic}, viz. in modelling isomerization, reactions with symmetric potential barriers\cite{Hazra}, in study of inversion of ammonia, in NV Centers \cite{NV_Centre} to name a few. Our interest in this model stems from our inclination in probing quantum tunneling with a simple system.

     As illustrated above,the propagator can be obtained analytically if and only if the classical action can be obtained in closed form. In this case, even though the Initial value problem (IVP) can be solved exactly \cite{Anharmonic1,Anharmonic4}, the boundary value problem is not solvable in closed form \footnote{The BVP solution is not even unique}. That restricts our venture of obtaining a closed form expression of classical action ($S^{cl}(x_t,x_0;t)$) in the perturbative regime, even though our method of computing the Path Integral does not inherently demand it.
     
     It might be noteworthy, unlike the Schulman\cite{Schulman} treatment, our analysis is not asymptotic and we do not neglect any contribution from paths. The perturbation theory arises solely due to the classical requirement and calculation of the quantum mechanical propagator for each order of $\lambda$ is exact. The exact solution can be obtained as (\ref{Propagator}), if the $S^{cl}(x_t,x_0;t)$ is instead solved for numerically.
     
    Consider the potential,
    
    \begin{align}\label{Pot}
        V(x)= \frac{1}{2}Max^2+\frac{1}{4}M \lambda Kx^4
    \end{align}

    where $\lambda$ is a small and positive parameter, $a<0, K>0$.  We compute the classical path using the Poincare-Linsted ansatz \cite{Landau1960classical} of perturbation theory,  upto the first order.
    
    \begin{align}\label{Path}
        &x_{cl}(t)=A \cos(\omega t+\phi)-\frac{\lambda K A^3}{8\omega_0^2}\big(\cos(\omega t+3\phi)\sin(\omega t)^2\big)
        \\ &\omega=\omega_0+\frac{3\lambda A^2}{8\omega_0}
    \end{align}

    The classical action of the system is obtained , by substituting the classical path, into the Lagrangian, integrating out time, and neglecting the terms of higher orders in $\lambda$. The resulting action is denoted by $S^{cl}(x_f,t;x_i,0)$ for the rest of the discussion. 
    The integration constants, A and $\phi$ , can be solved for, using the boundary conditions $x(0)=x_i,\;x(t)=x_f$ and are substituted in expression of the classical action. The inverse solution is not unique, and we choose the solution in which both A,$\phi$ are real and positive.
    %The details of the calculation can be found in Appendix.
    \begin{align}
        &A=\sqrt{ x_f ^2+ x_i ^2-2 x_f x_i \cos[t \omega_0]} \csc[t \omega_0]
        \\&\phi  =\arccos\left[\frac{ x_i \sin[t \omega_0]}{\sqrt{ x_f ^2+ x_i^2-2 x_f x_i\cos[t \omega_0]}}\right]
    \end{align}
    
    \par The perturbative proagator  is given by,

        \begin{multline}\label{Propagator}
        K(x_b,x_a; t, 0 )=\\
        \sqrt{\frac{M }{2\pi i \hbar t }} \exp{\frac{i S^{cl}(x_b,x_a; t)}{\hbar}}\sum_{k=0}^{\infty}\frac{1}{k!} \Big(\frac{i \hbar \lambda K}{ 4  M}\Big)^k \Big[ \int_0^t \frac{1}{\omega(t')^2}dt' \Big]^{k}  \\
        \times\Big(\frac{\partial}{\partial \phi}\Big)^{2k} \sqrt{\frac{\phi}{\sin{(\phi)}}}  
        \end{multline}
        
    \par The Angular frequency is defined as,

    \begin{align}
        \omega(t)=\sqrt{\frac{1}{M}\frac{\partial^2 V(x)}{\partial x^2}|_{x=x(t)}}
    \end{align}

    Substituting (\ref{Pot}), neglecting higher order terms and using the formula of double-angles, the angular frequency becomes,

    \begin{align}
        \omega(t)=\sqrt{a + \frac{3K \lambda A^2}{2}+\frac{3K \lambda A^2}{2} \cos[2(\omega_0+\frac{3\lambda A^2}{8\omega_0}) t+2\phi]}
    \end{align}
     
    The integral involving $\omega(t)$ in (\ref{Propagator}) can be computed analytically \cite{Gradshteyn,Byrd},

\begin{widetext}
  
      \begin{align}
       &\int_0^{t} \omega(t')^{-2} \; dt'=\frac{\arctan\left[\frac{w_0^2\tan\left[\frac{1}{2} t \left(\frac{3 A^2 K \lambda }{8 w_0}+w_0\right)\right]}{\sqrt{-\frac{9}{4}A^4 K^2 \lambda ^2+\left(\frac{3}{2} A^2 K \lambda+w_0^2\right){}^2}}\right]}{\left(\frac{3 A^2 K\lambda }{8 w_0}+w_0\right) \sqrt{-\frac{9}{4}A^4 K^2 \lambda ^2+\left(\frac{3}{2} A^2 K \lambda +w_0^2\right){}^2}}
    \end{align}
    
\begin{figure}[t]
\begin{subfigure}{0.49\textwidth}
\includegraphics[width=\linewidth]{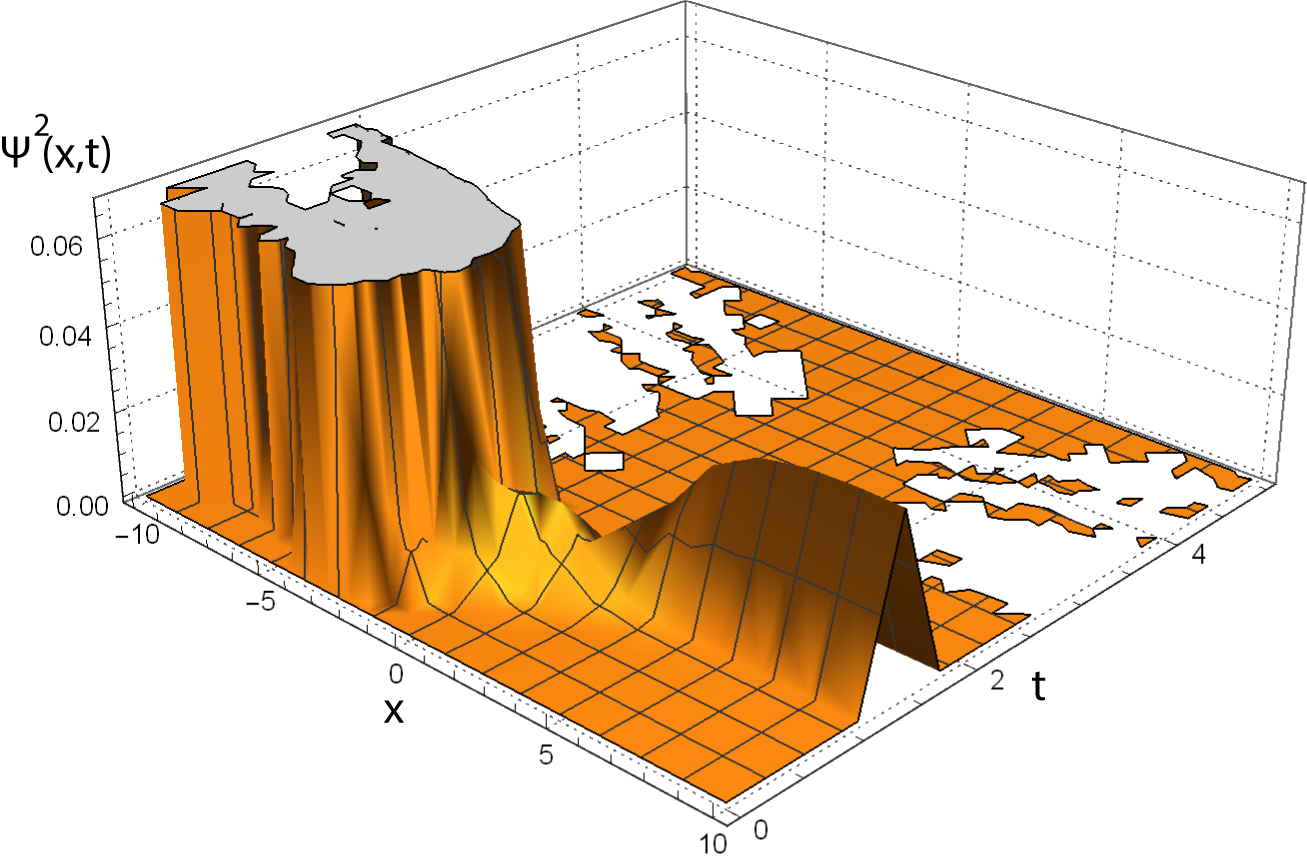} 
\caption{Probability density as a function of x,t}
\label{fig:subim1}
\end{subfigure}
\begin{subfigure}{0.49\textwidth}
\includegraphics[width=\linewidth]{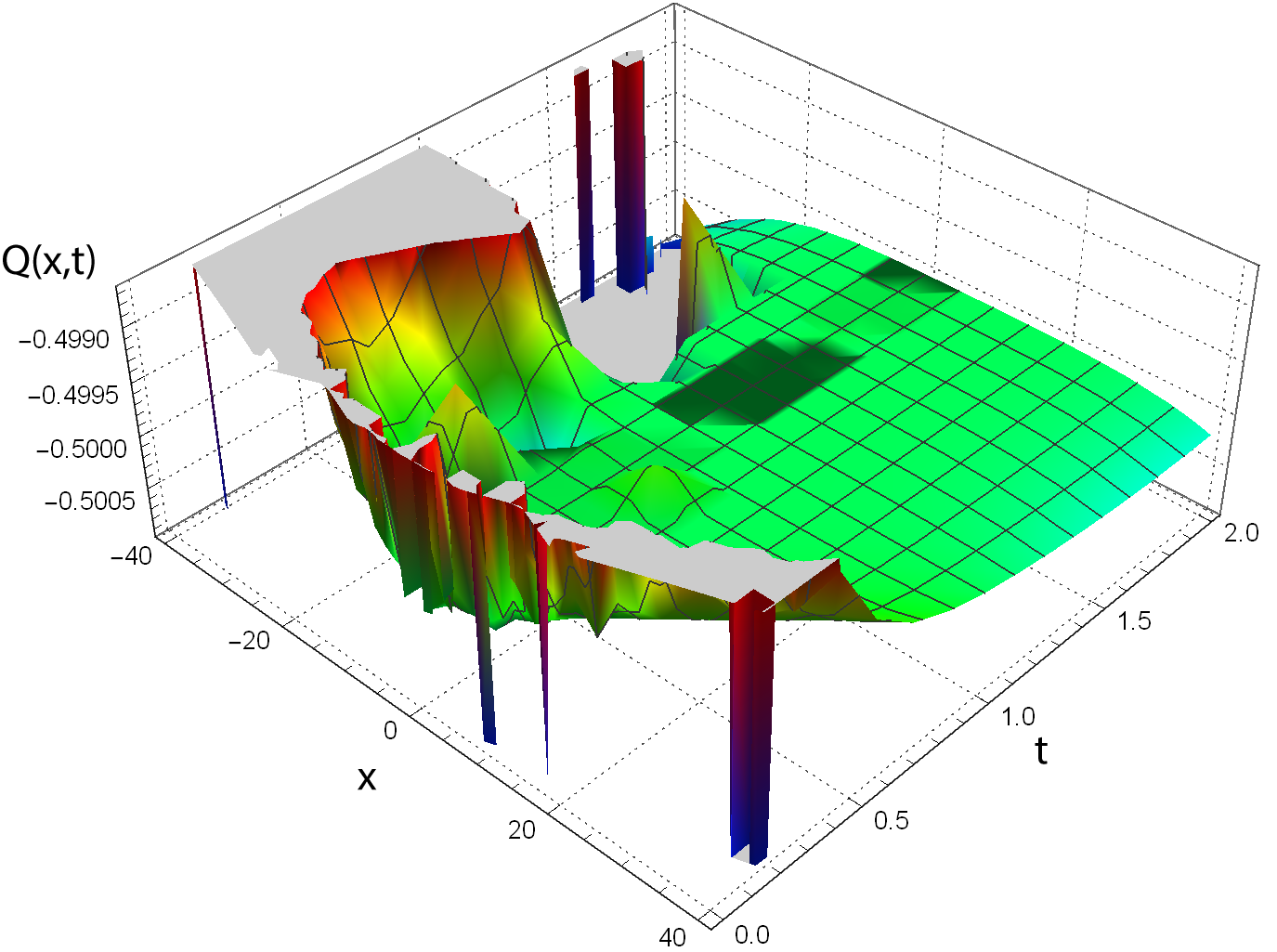}
\caption{Quantum potential as a function of x,t}
\label{fig:subim2}
\end{subfigure}

\caption{Plots for the Probability Density and Quantum Potential for the double-well with $M=1,a=-2,\lambda=10^{-4}, K=1,\hbar=1$} 
\label{fig:image2}

\end{figure}

    \end{widetext}

    The figure (\ref{fig:image2}) reports the probability density and the Quantum Potential as function of space and time, computed numerically using equations (\ref{dynamics}), for the parameter values $M=1,a=-2,\lambda=10^{-4}, K=1,\hbar=1$ and the initial Gaussian wave packet,
    
    \begin{align}
        \Psi_0(x_i)=\frac{1}{\sqrt{2 \pi \alpha^2}} e^{\frac{-(x_i - l)^2}{2 \alpha^2}}
    \end{align}
    where for the purpose of the plot we have chosen $\alpha=0.4$ and $l=-3.126$.
    The parameters of the initial wave function is chosen such that the initial probability of finding the particle in the left well is $\sim1$, as well as it is centered very close to the barrier wall to show some interesting dynamics, even in its initial time of flight.

    \par As is expected from earlier studies, a part of the probability density is clearly observed to cross the classically forbidden barrier, which is situated at $x=0$. The QP is shown to develop a deep groove in the left well along with a smaller one in the right one. It is the sum of QP and the applied potential V(x), that governs the dynamics of the quantum fluid. 
    
    \par Bohm had interpreted the mechanism of tunneling by proposing that the Quantum Potential creates channels, by lowering the applied potential. If the QP is negative then the effective total potential is lower than the actual applied potential. Then the whole dynamics can be seen as a blob of fluid traversing through the lowered potential.\cite{Bohm1} On the other hand according to Wyatt \cite{Wyatt1} et al, it is the initial position dependent acceleration, during initial period of flight, that causes certain fluid elements to fly over the classical barrier. That is, whether a particle will be tunneled or not, is encoded in its initial position in the wave-packet which determines the acceleration for initial moments. This in turn would dictate whether the particle will finally gather enough momentum to cross the barrier.
    
    \par Our results indicate a reconciliation of the two seemingly contradicting views. Bohm's interpretation is validated from the flat groove travelling towards the right well in the Quantum Potential. On the other hand the QP is steep initially, and thus its gradient is large. Depending on the initial position of the fluid element in the well, the force either directs it to the left or the right well during the time of flight. With time the QP becomes flat and thus its gradient becomes zero. So, it is only the initial time of flight that decides the fate of the fluid element in crossing the barrier, and to a good approximation the QP can be neglected for further dynamics, as was conceived by Wyatt. Both the interpretations are thus found to be completely consistent with each other as far as this toy model is concerned \footnote{Other studies on tunneling using quantum trajectories can be found in \cite{Streamlines}}. Exploration in this direction using more complicated and general models forms material for further study.

    In a double well potential the Probability amplitude is expected to oscillate between two wells, which is not evident in our plots, and probably is rendered due to higher order corrections. The higher order effects were not incorporated as our main interest was to illustrate tunneling, which is well studied even within first order. Their incorporation is straight forward numerically.

\section{Discussion}
     \par QP is regarded as the origin of non-locality in QFD\cite{Bohm1,Wyatt1}. Invoking Feynman's idea of contribution from all possible paths in governing the dynamics, and quantifying their overall effect the formulation presented in this paper, unravels a new arena of possibilities in the understanding of the processes quantum nonlocality and entanglement, which are of utmost importance in the fields of Quantum Meteorology, Condensed Matter systems and most importantly in developing Quantum Computers. The various terms in the series in (\ref{GeneralIntegral}), quantify different kinds of correlations, over the classical action. This is best explained in an analogy with Quantum Circuits.
     
     \par Consider the space to be a discrete lattice. In the circuit model of quantum computation a finite dimensional Hilbert space (usually qubits) lives at every lattice points, and the time evolution is given by unitary gates connecting qubits of an instant with the next. The number of qubits one gate mixes by its \textit{action}, is given by the number of legs (input and output) that the gate has. Different legged gates thus generates different class of correlations. Consider a discrete version of (\ref{GeneralIntegral}). The integrals are changed into sums, and various order of $k$ generates correlations between various modes of $m$, which in turn generates spatial and temporal correlations through the various partial derivatives of the potential. Likewise the series allows one to study separate orders of correlation w.r.t the classical action for continuum systems separately. 
     
     \par It might be worthwhile to note, we have not used any extra assumptions than the original quantum theories and have merely mathematically connected results of two preexisting seemingly disconnected nodes. Recently, Quantum Fluids have found recent applications in the non-linear framework of Gross-Pitaevski equations\cite{Askar2018}, which governs the dynamics of BEC, soliton-polariton semiconductor systems etc. There are several straight-forward scopes of generalizing our work to these domains, to generate analytical solutions. From a mathematical perspective, this problem is interesting, as the integral formulation corresponding to a non-linear PDE is not well understood. Enquiry in this direction can open up an opportunity to study at least a special class of those.

\section{Concluding Remarks}

     \par To summarize, in this paper we have bridged the Quantum Fluid Dynamics with the Path Integral formulation and presented a general route to the derivation of QP and Quantum Trajectories from the principles of Path Integrals, expressing it analytically as a functional of the classical path and the initial wave-function of a system. These analytical expressions are valid for any general well-behaved potential, and can be provided to computers directly as inputs, completely bypassing the task of solving the Schrodinger's equation. For any given initial wave-function the computation requires to solve the boundary value problem for the classical path thus making it faster than currently available algorithms.

\begin{acknowledgments}
     The present paper reports a part of the work done at Homi Bhabha Centre for Science Education (Tata Institute of Fundamental Research), Mumbai. We are indebted to the National Initiative on Undergraduate Science (NIUS), Chemistry fellowship of HBCSE (Batch XIII, 2016-2018) for their tremendous and inspiring support. We acknowledge the support of the Govt. Of India, Department of Atomic Energy, under Project No. 12-R\&D-TFR-6.04-0600. Many of the colleagues at the programme and the institute have enriched us. SG wishes to thank Dr. Anirban Hazra for several stimulating discussions. No language of acknowledgement is enough for the utmost care that Dr. Amrita B. Hazra has provided to SG. Without the thorough support she has offered throughout, it would have not been possible to conceive the project to its present extent.
\end{acknowledgments}

\bibliography{QFD}

\end{document}